\documentclass[fleqn,usenatbib]{mnras}

\usepackage[T1]{fontenc}
\usepackage{ae,aecompl}
\usepackage[normalem]{ulem}

\usepackage{newtxtext,newtxmath}

\usepackage{graphicx}	
\usepackage{amsmath}	
\usepackage{amssymb}	
\usepackage[shortcuts]{extdash}
\usepackage{calc}
\usepackage{multirow}

\newcommand{\kcrb}{$\kappa$\,CrB}
\newcommand{\lyn}{6\,Lyn}
\newcommand{\sex}{24\,Sex}
\newcommand{\hdone}{HR\,6817}
\newcommand{\hdtwo}{HR\,8461}
\newcommand{\teff}{$T_\mathrm{eff}$}
\newcommand{\logg}{$\log(g)$}
\newcommand{\feh}{[Fe/H]}
\newcommand{\msun}{$\mathrm{M}_\odot$}

\newcommand{\um}{$\mu$m}
\newcommand{\fbol}{$F_{\mathrm{bol}}$}
\newlength{\okinalen}
\newcommand{\okina}{\hbox to .666\okinalen{\hss`\hss}}

\title[Interferometric diameters of retired A stars]{Interferometric diameters of five evolved intermediate-mass planet-hosting stars measured with PAVO at the CHARA Array}

\author[T. R. White et al.]
{T.~R.~White,$^{1,2,3}$\thanks{E-mail: tim@phys.au.dk}
D.~Huber,$^{4,5,1,6}$ 
A.~W.~Mann,$^{7,8,9}$ 
L.~Casagrande,$^{10,11}$ 
S.~K.~Grunblatt,$^{4}$ 
\newauthor
A.~B.~Justesen,$^{1}$ 
V.~Silva~Aguirre,$^{1}$
T.~R.~Bedding,$^{5,1}$
M.~J.~Ireland,$^{10}$
G.~H.~Schaefer,$^{12}$
\newauthor
P.~G.~Tuthill$^{5}$
\\
$^{1}$Stellar Astrophysics Centre, Department of Physics and Astronomy, Aarhus University, Ny Munkegade 120, DK-8000 Aarhus C, Denmark\\
$^{2}$Institut f\"{u}r Astrophysik, Georg-August-Universit\"{a}t G\"{o}ttingen, Friedrich-Hund-Platz 1, 37077 G\"{o}ttingen, Germany\\
$^{3}$Max-Planck-Institut f\"ur Sonnensystemforschung, Justus-von-Liebig-Weg 3, 37077 G\"ottingen, Germany\\
$^{4}$Institute for Astronomy, University of Hawai\okina i, 2680 Woodlawn Drive, Honolulu, HI 96822, USA\\
$^{5}$Sydney Institute for Astronomy (SIfA), School of Physics, University of Sydney, NSW 2006, Australia\\
$^{6}$SETI Institute, 189 Bernardo Avenue, Mountain View, CA 94043, USA\\
$^{7}$Department of Astronomy, The University of Texas at Austin, Austin, TX 78712, USA \\
$^{8}$Department of Astronomy, Columbia University, 550 West 120th Street, New York, NY 10027, USA \\
$^{9}$NASA Hubble Fellow\\
$^{10}$Research School of Astronomy and Astrophysics, Mount Stromlo Observatory, The Australian National University, ACT 2611, Australia\\
$^{11}$ARC Centre of Excellence for All Sky Astrophysics in 3 Dimensions (ASTRO 3D)\\
$^{12}$The CHARA Array of Georgia State University, Mount Wilson Observatory, Mount Wilson, CA 91023, USA
}

\date{Accepted XXX. Received YYY; in original form ZZZ}

\pubyear{2018}

\begin{document}
\label{firstpage}
\pagerange{\pageref{firstpage}--\pageref{lastpage}}
\maketitle

\begin{abstract}
Debate over the planet occurrence rates around intermediate-mass stars has hinged on the accurate determination of masses of evolved stars, and has been exacerbated by a paucity of reliable, directly-measured fundamental properties for these stars. We present long-baseline optical interferometry of five evolved intermediate-mass ($\sim$\,1.5\,\msun) planet-hosting stars using the PAVO beam combiner at the CHARA Array, which we combine with bolometric flux measurements and parallaxes to determine their radii and effective temperatures. We measured the radii and effective temperatures of 6\,Lyncis (5.12$\pm$0.16\,$\mathrm{R}_\odot$, 4949$\pm$58\,K), 24\,Sextantis (5.49$\pm$0.18\,$\mathrm{R}_\odot$, 4908$\pm$65\,K), $\kappa$\,Coronae\,Borealis (4.77$\pm$0.07\,$\mathrm{R}_\odot$, 4870$\pm$47\,K), HR\,6817 (4.45$\pm$0.08\,$\mathrm{R}_\odot$, 5013$\pm$59\,K), and HR\,8641 (4.91$\pm$0.12\,$\mathrm{R}_\odot$, 4950$\pm$68\,K). We find disagreements of typically 15\,per cent in angular diameter and $\sim$\,200\,K in temperature compared to interferometric measurements in the literature, yet good agreement with spectroscopic and photometric temperatures, concluding that the previous interferometric measurements may have been affected by systematic errors exceeding their formal uncertainties. Modelling based on BaSTI isochrones using various sets of asteroseismic, spectroscopic, and interferometric constraints tends to favour slightly ($\sim$\,15\,per cent) lower masses than generally reported in the literature.
\end{abstract}

\begin{keywords}
stars: fundamental parameters -- techniques: interferometric -- planetary systems
\end{keywords}



\section{Introduction}

Planet occurrence rates as a function of host star properties are of key interest to interpret exoplanet demographics and constrain planet formation scenarios. In particular, a correlation between gas-giant planet occurrence and stellar mass \citep{johnson10}, and the preference for small planets around cooler stars \citep{latham10,howard10} have been interpreted as evidence for the core-accretion scenario as the dominant mechanism of planet formation. However, traditional planet detection methods, such as radial velocities and transits, become insensitive for intermediate-mass main-sequence stars due to rapid rotation and pulsations. While pulsation timings have recently been used to detect a planet around a main-sequence A star \citep{murphy16}, and several exoplanets transiting main-sequence A stars have now been discovered \citep[e.g.][]{CollierCameron10,Hartman15,Morton16,Zhou16,Gaudi17}, the majority of constraints for planet occurrence rates in intermediate-mass stars still rely on Doppler searches around evolved G- and K-type subgiants and giants, sometimes refered to as ``retired A stars'' \citep{frink02,sato03,hatzes03,johnson07a,niedzielski07}.

The difficulty of measuring masses for such evolved stars from spectroscopy and stellar isochrones has led to a debate over the reality of the correlation between planet occurrence and stellar mass for gas-giant planets \citep[e.g.][]{lloyd11,johnson13,schlaufman13}. While recent studies focused on asteroseismology as a way to independently test ``spectroscopic'' masses \citep{ghezzi15, campante17, north17, Stello17}, accurate effective temperatures and radii from long-baseline interferometry play a key role for resolving the model-dependent systematic errors. A number of bright intermediate-mass giants have both detected solar-like oscillations and interferometric measurements, including Pollux \citep{Mozurkewich03,hatzes12}, $\iota$\,Dra \citep{Zechmeister08,Baines11}, $\xi$\,Hya \citep{Frandsen02,Thevenin05}, $\epsilon$\,Oph \citep{Barban07,Mazumdar09}, and HD\,185351 \citep{Johnson14}, allowing for mass to be inferred independent from spectroscopy. In the case of HD\,185351, the extra asteroseismic information provided by \emph{Kepler} photometric measurements, in conjunction with interferometry and high-resolution spectroscopy has allowed for excellent tests of stellar evolutionary models \citep{Hjoerringgaard17}. Additional examples of evolved planet hosts are required to test whether these results are systematic.

\begin{table*}
\centering
\caption{Stellar properties from the literature}
\label{tab:spect}
\begin{tabular}{lcccccccc}
\hline
Star     & HD     & Sp. type & $T_\mathrm{eff}$ & $\log(g)$     & [Fe/H]           & Mass & Ref. & Parallax$^a$        \\
         &        &          & (K)              & (dex)         & (dex)            & M$_\odot$ &     & (mas)           \\
\hline
\lyn     &  45410 & K0III-IV & 4938$\pm$25      & 3.19$\pm$0.03 & $+$0.01$\pm$0.01 & 1.44$\pm$0.14 & \citet{Brewer16}  &  17.92$\pm$0.47 \\
\sex     &  90043 & G5IV     & 5069$\pm$62      & 3.40$\pm$0.13 & $-$0.01$\pm$0.05 & 1.81$\pm$0.08 & \citet{Mortier13} &  12.91$\pm$0.38 \\
\kcrb    & 142091 & K0III-IV & 4876$\pm$46      & 3.15$\pm$0.14 & $+$0.13$\pm$0.03 & 1.58$\pm$0.08 & \citet{Mortier13} &  32.79$\pm$0.21 \\
HR\,6817 & 167042 & K1III    & 5028$\pm$53      & 3.35$\pm$0.18 & $+$0.03$\pm$0.04 & 1.63$\pm$0.06 & \citet{Mortier13} &  19.91$\pm$0.26 \\
HR\,8461 & 210702 & K1III    & 5000$\pm$44      & 3.36$\pm$0.08 & $+$0.04$\pm$0.03 & 1.71$\pm$0.06 & \citet{Mortier13} &  18.20$\pm$0.39 \\
\hline
\end{tabular}
\newline \flushleft $^a$ Parallax values from \citet{vanleeuwen07}.
\end{table*}

A related debate surrounds the accuracy of interferometric angular diameters themselves. While interferometry is often considered as the ``ground-truth'', it is important to realize that interferometric visibilities can be affected by strong systematic errors due assumed calibrator sizes, wavelength scales, and limb-darkening corrections. Such differences can have a significant impact on the calibration of effective temperatures scales. For example, systematic differences between photometric temperatures from the infrared flux method and CHARA $K'$-band diameters have been noted for angular sizes $\lesssim$\,1\,mas \citep{Casagrande14}, and smaller diameters measured in $H$ band showed better agreement \citep{huang15b}. Since calibration errors are more severe for smaller angular diameters (corresponding to more unresolved sources, given a fixed baseline and wavelength), this indicates that some diameters measured with long baseline optical interferometry may be affected by systematic errors. Understanding (and correcting) such systematic errors is critical to establishing fundamental temperature scales, and thus also to settling the debate over the masses of evolved stars. Astrophysical phenomena, including starspots \citep[e.g][]{roettenbacher16,richichi17} and unresolved companions, may also affect angular diameter measurements.

A solar metallicity F0 star has a main-sequence mass of $\sim$1.55\,M$_\odot$, while an A star has a (model- and metallicity-dependent) mass range of $\sim$1.6--2.4\,M$_\odot$ \citep{gray92}. Rather than focusing on the semantics of this definition, in this paper we will simply focus on measurements of stars that have been included in samples of so-called retired A stars. We present optical long-baseline interferometry of five suspected retired A stars (\lyn, \sex, \kcrb, \hdone, and \hdtwo) to measure accurate effective temperatures and radii and explore systematic errors in interferometric angular diameters. Additionally, we present model-dependent masses derived from various sets of interferometric, spectroscopic, and asteroseismic constraints. 

Each of our targets hosts a confirmed exoplanet as part of the original retired A star sample \citep{johnson07a,Johnson08,Sato08,Johnson11}. Properties of the stars from the literature are given in Table~\ref{tab:spect}. Four of the targets also have previously published interferometric angular diameters, mostly using near-infrared measurements from the Classic beam combiner at the CHARA Array \citep{Baines09,Baines10,vonBraun14}, but also with measurements at visible wavelengths with the VEGA beam combiner at the CHARA array \citep{Ligi16} and NPOI \citep{Baines13}. \citet{Stello17} have recently presented asteroseismic detections for three of these targets, amongst others, using the Hertzsprung SONG telescope \citep{Grundahl17}.

\section{Observations}

\subsection{PAVO interferometry}
We made interferometric observations with the PAVO beam combiner \citep{Ireland08} at the CHARA Array at Mount Wilson Observatory, California \citep{tenBrummelaar05}. The CHARA Array consists of six 1-m telescopes in a Y-shaped configuration, with baselines ranging from 34 to 331\,m. PAVO, one of several beam combiners operating at CHARA, is a pupil-plane combiner that operates at visible wavelengths ($\sim$600--800\,nm), with a limiting magnitude in typical seeing conditions of $R\sim8$\,mag. PAVO may combine light from two or three telescopes, although calibration of the fringe visibilities is more robust when operating in two-telescope mode.

Observations were made over several observing seasons; a summary is given in Table~\ref{tab:obs}. Instrumental and atmospheric effects combine to cause raw fringe visibility measurements to be significantly lower than the true visibility, necessitating calibration. To do this, calibration stars with reasonably well-known sizes are observed. To minimize the impact of errors in calibrator diameter sizes, these calibrator stars need to be as unresolved by the interferometer as possible, and several calibrator stars are used for each target. Additionally, to minimize the effects of spatial and temporal variations in the system visibility, they should be observed as closely as possible to the target, that is within 10$^\circ$, and immediately before and after a observation of a target. Observations were conducted in the sequence \emph{calibrator 1} -- \emph{target} -- \emph{calibrator 2}, with two minutes of visibility measurements made for each star. Including slewing, such a sequence typically takes 15\,min.

The list of the calibrator stars we have used is given in Table~\ref{tab:cals}. Calibrator angular diameter sizes were estimated from the ($V-K$) surface brightness relation of \citet{boyajian14}. Magnitudes in $V$ band were taken from the Tycho-2 catalogue \citep{Hoeg2000} and converted into the Johnson system using the calibration by \citet{Bessell00}, while those in $K$ band were taken from the 2MASS catalogue \citep{Skrutskie2006}. Reddening was estimated using the dust map of \citet{Green15}. Finally, the diameters were corrected for limb darkening to determine their corresponding $R$-band uniform disc diameter.

Raw observations were reduced to produce calibrated visibilities using the PAVO reduction software, which has been well-tested and used for multiple studies \citep[e.g.][]{Bazot11,derekas11,huber12,maestro13}.

\begin{table}
\centering
\caption{Log of PAVO interferometric observations}
\label{tab:obs}
\begin{tabular}{lcccc}
\hline
\textsc{UT} Date & Baseline$^a$ & Target & No. scans & Cal.$^b$\\
\hline
2012 Sept 7 & W1W2 & \hdtwo & 3 & mno \\
            & S2W2 & \lyn   & 2 & bc  \\
2013 July 6 & E2W2 & \hdtwo & 3 & mno \\
2013 July 7 & W1W2 & \hdone & 3 & jk \\
2013 July 8 & E2W2 & \hdone & 3 & jkl \\
            &      & \hdtwo & 3 & mno \\
2013 July 9 & E2W2 & \hdone & 3 & jk \\
2014 Feb 21 & E2W2 & \sex   & 2 & ef \\
2014 Apr 6  & W1W2 & \kcrb  & 1 & gh \\
2014 Apr 8  & E2W1 & \sex   & 3 & ef \\
2014 Nov 9  & S1W2 & \lyn   & 4 & bd \\
2014 Nov 10 & E2W2 & \lyn   & 4 & ad \\
2015 Apr 4  & W1W2 & \kcrb  & 5 & i \\
\hline
\end{tabular}
\newline \flushleft $^a$ The baselines used have the following lengths: W1W2, 107.92\,m; E2W2, 156.27\,m; S1W2, 210.97\,m.
\newline $^b$ Refer to Table~\ref{tab:cals} for details of the calibrators used.
\end{table}

\begin{table}
\centering
\caption{Calibration stars used for observations}
\label{tab:cals}
\begin{tabular}{lcccccc}
\hline
HD & Sp. type & $V$ & $K$ & $E(B-V)$ & $\theta_{\mathrm{UD},R}$ & ID \\
   &          & (mag) & (mag) & (mag) & (mas) & \\
\hline
 38129 & A0     & 6.795 & 6.440 & 0.114 & 0.173(9) & a \\
 40626 & B9.5IV & 6.043 & 6.106 & 0.016 & 0.195(10) & b \\
 46294 & A0     & 6.840 & 6.594 & 0.022 & 0.163(8) & c \\
 46590 & A2V    & 5.873 & 5.799 & 0.008 & 0.230(12) & d \\
 85504 & A0III/IV & 6.015 & 6.020 & 0.015 & 0.205(10) & e \\
 90763 & A0V    & 6.041 & 5.937 & 0.005 & 0.217(11) & f \\
138341 & A4IV   & 6.456 & 5.810 & 0.006 & 0.248(12) & g \\
144206 & B9III  & 4.720 & 4.880 & 0.004 & 0.341(17) & h \\
144359 & A0     & 6.774 & 6.481 & 0.011 & 0.172(9) & i \\
161693 & A2V    & 5.751 & 5.585 & 0.017 & 0.257(13) & j \\
169885 & A3m    & 6.352 & 5.955 & 0.005 & 0.223(11) & k \\
173664 & A2IV   & 6.194 & 5.806 & 0.009 & 0.238(12) & l \\
208108 & A0Vs   & 5.680 & 5.631 & 0.009 & 0.248(12) & m \\
209459 & B9.5V  & 5.828 & 5.882 & 0.043 & 0.216(11) & n \\
214203 & A1III  & 6.428 & 6.336 & 0.016 & 0.180(9) & o \\
\hline
\end{tabular}
\end{table}

\subsection{Spectrophotometry and bolometric fluxes}
To determine interferometric effective temperatures, the measured angular diameters must be combined with a measurement of the bolometric flux at Earth (\fbol):
\begin{equation}
T_\mathrm{eff} = \left(\frac{4F_\mathrm{bol}}{\sigma\theta_\mathrm{LD}^2}\right)^{1/4},
\end{equation}
where $\sigma$ is the Stefan-Boltzmann constant, and $\theta_\mathrm{LD}$ is the measured angular diameter after correction for limb-darkening. 

To obtain \fbol\ measurements we acquired optical spectra for our targets with the SuperNova Integral Field Spectrograph \citep[SNIFS;][]{Aldering02,Lantz04}, operating at the University of Hawai\okina i 2.2-m telescope on Maunakea. SNIFS provides low-resolution ($R\simeq1000$) spectra between 320--970\,nm, with excellent spectrophotometric precision. 
All targets were observed on 2017 April 8 and 9 under clear conditions. Since these targets are quite bright, the SNR exceeded 400 around 6000\,\AA\ (per pixel) for each target. However, in this high-SNR regime, bolometric flux determinations are limited primarily by the spectrophotometric calibration \citep[1--2 per cent;][]{Mann13}.

Bolometric fluxes were computed by integrating over absolutely flux calibrated spectra, built primarily from our optical spectra and NIR templates from the Infrared Telescope Facility (IRTF) Cool Stars library \citep{Rayner2009}. For \kcrb\ we used an optical spectrum from Hubble's Next Generation Spectral Library \citep[NGSL,][]{Heap2007}, which is more precise and has better wavelength coverage than the SNIFS spectra. 

\begin{figure}
    \includegraphics[width=\columnwidth]{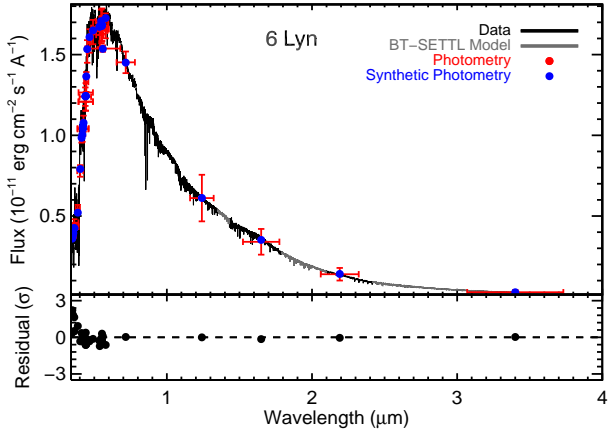}
    \caption{Flux-calibrated spectrum of \lyn, from which we compute \fbol. Black data shows the empirical spectra from SNIFS in the optical, and the IRTF library in the NIR. Grey regions indicate areas of high telluric contamination or beyond the reach of our empirical spectra, which we have filled in using atmospheric models. Red points are literature photometry, with error bars in the Y-axis indicating reported measurement uncertainties, and errors in the X-axis indicating the effective width of the filter. Synthetic photometry computed from the displayed spectrum is shown as blue points. Estimated residuals (observed - synthetic photometry) are shown in the bottom panel in units of standard deviations. Equivalent figures for the other stars in our sample are provided in Appendix~\ref{apx:fbol}.
    }
    \label{fig:fbol}
\end{figure}

We joined and calibrated the optical and NIR spectra following the procedure from \citet{Mann2015b}, which we briefly summarize here. For each target, we downloaded published optical and NIR photometry from the Two-Micron All-Sky Survey \citep[2MASS,][]{Skrutskie2006}, Tycho-2 \citep{Hoeg2000}, Hipparcos \citep{vanLeeuwen97}, the Wide-field Infrared Survey Explorer \citep[WISE,][]{Wright2010}, and The General Catalogue of Photometric Data \citep[GCPD,][]{Mermilliod1997}. We computed synthetic magnitudes from each spectrum using the appropriate filter profile and zero-point \citep{cohen03,Jarrett2011,Mann2015a}. We replaced regions of high telluric contamination and those not covered by our spectra (e.g., beyond 2.4\um) with a best-fit atmospheric model from the BT-SETTL grid \citep{Allard2011}. The spectra were scaled to match the photometry, using the overlapping NIR and optical spectra (0.8-0.95\um) as an additional constraint. We show an example calibrated spectrum in Fig.~\ref{fig:fbol}.

Uncertainties were computed by repeating the process for each star, varying input parameters with random and correlated errors (e.g., flux calibration, filter zero-points and profiles), then recomputing \fbol\ each time. Uncertainties in the zero-points and filter profiles for Tycho and Hipparcos photometry amount to about 2 per cent, with similar zero-point uncertainties for the other photometry.  The Hipparcos and Tycho calibration, built on STIS spectra from NGSL \citep{Heap2007}, is accurate to 0.5 per cent \citep{Mann2015a}.

Except for \kcrb, which has a NIR spectrum in the IRTF library, we used other IRTF library templates of similar spectral type to approximate the true NIR spectrum. We also explored uncertainties due to template choice by re-joining the spectra and computing \fbol\ with any template from the IRTF library within two spectral spectral subtypes of the target. Resulting uncertainties in \fbol\ are generally small (2--5 per cent), owing to the wealth of optical photometry available for these stars, and the comparatively low flux in the NIR, where the spectral shape is most uncertain. 

We have additionally derived estimates for \fbol\ from bolometric corrections determined from MARCS model atmospheres fluxes \citep{Gustafsson08} by \citet{Casagrande18}. We used the bolometric corrections for Hipparcos \citep{Hipparcos97} and Tycho2 photometry \citep{Hoeg2000} at the spectroscopic $T_\mathrm{eff}$, $\log g$, and [Fe/H] given in Table~\ref{tab:spect}. We assumed zero reddening because the stars are all nearby. Uncertainties were computed from MonteCarlo simulations with the reported spectroscopic and photometric uncertainties. These uncertainties do not account for possible deficiencies in MARCS synthetic fluxes, but extensive comparisons with observations usually validate them at the level of a few per cent \citep{Casagrande18}. The final bolometric flux, denoted $F_\mathrm{bol,MARCS}$ in Table~\ref{tab:newresults}, was determined from a weighted average across the values from the Hipparcos, and Tycho $B_T$ and $V_T$ magnitudes.

While we adopted the bolometric fluxes as determined above from spectrophotometry ($F_\mathrm{bol,sp}$) for our final values here because they have less model dependence, the MARCS fluxes may be applied more readily for other stars, and so it is instructive to see how well they compare. We generally find excellent agreement, except for \hdone, for which $F_\mathrm{bol,MARCS}$ is smaller by 2.7$\sigma$.

An alternative estimation of \fbol\ may be made from a calibration of Tycho2 photometry, which has been built upon a sample of stars for which the infrared flux method (IRFM) has been applied \citep{Casagrande06,Casagrande10}. This sample is dominated by main-sequence stars, and the few giants in the sample also tend to be metal-poor. Additionally, the relations have a dependence on magnitude, and bright stars were saturated in the calibration sample. For these reasons, we do not expect the bolometric fluxes determined from this method for our targets to be accurate to better than a few percent. Indeed, for this method we find values are, on average, 3.5 per cent larger than than the $F_\mathrm{bol,sp}$ measurements, although with only five stars this difference is not statistically significant.

\section{Results}

\begin{figure}
    \includegraphics[width=\columnwidth]{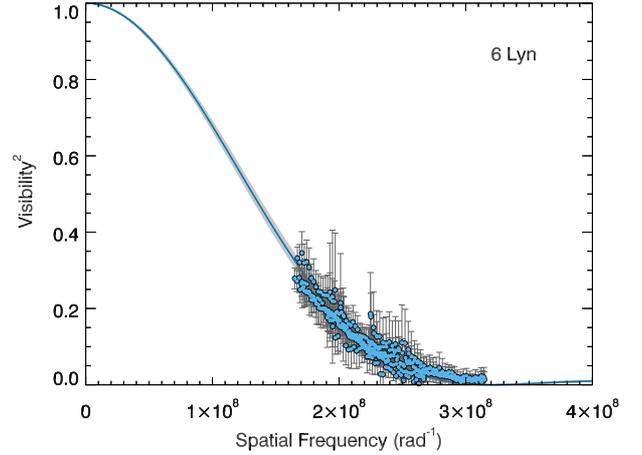}
    \caption{Squared visibility versus spatial frequency for \lyn. The blue line shows the fitted limb-darkened model to the PAVO observations (blue circles), with the light grey-shaded region indicating the 1-$\sigma$ uncertainties. Note that the error bars have been scaled so that the reduced $\chi^{2}$ equals unity. 
    }
    \label{fig:6Lyn}
\end{figure}

\begin{figure}
	\includegraphics[width=\columnwidth]{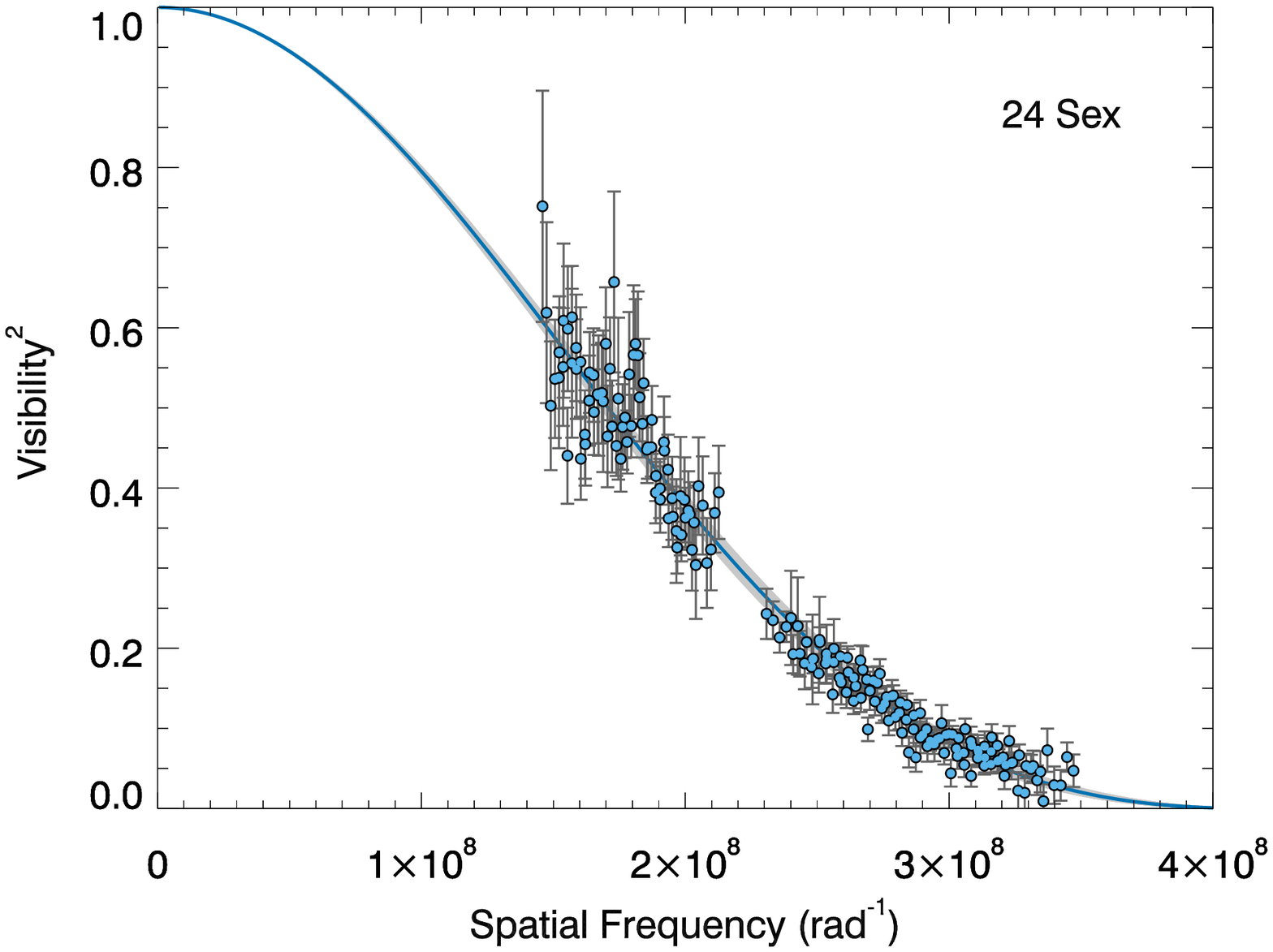}
    \caption{Squared visibility versus spatial frequency for \sex. The blue circles and line indicate the PAVO observations and best-fitting model, respectively, as for Fig.~\ref{fig:6Lyn}.}
    \label{fig:24Sex}
\end{figure}

\begin{figure}
	\includegraphics[width=\columnwidth]{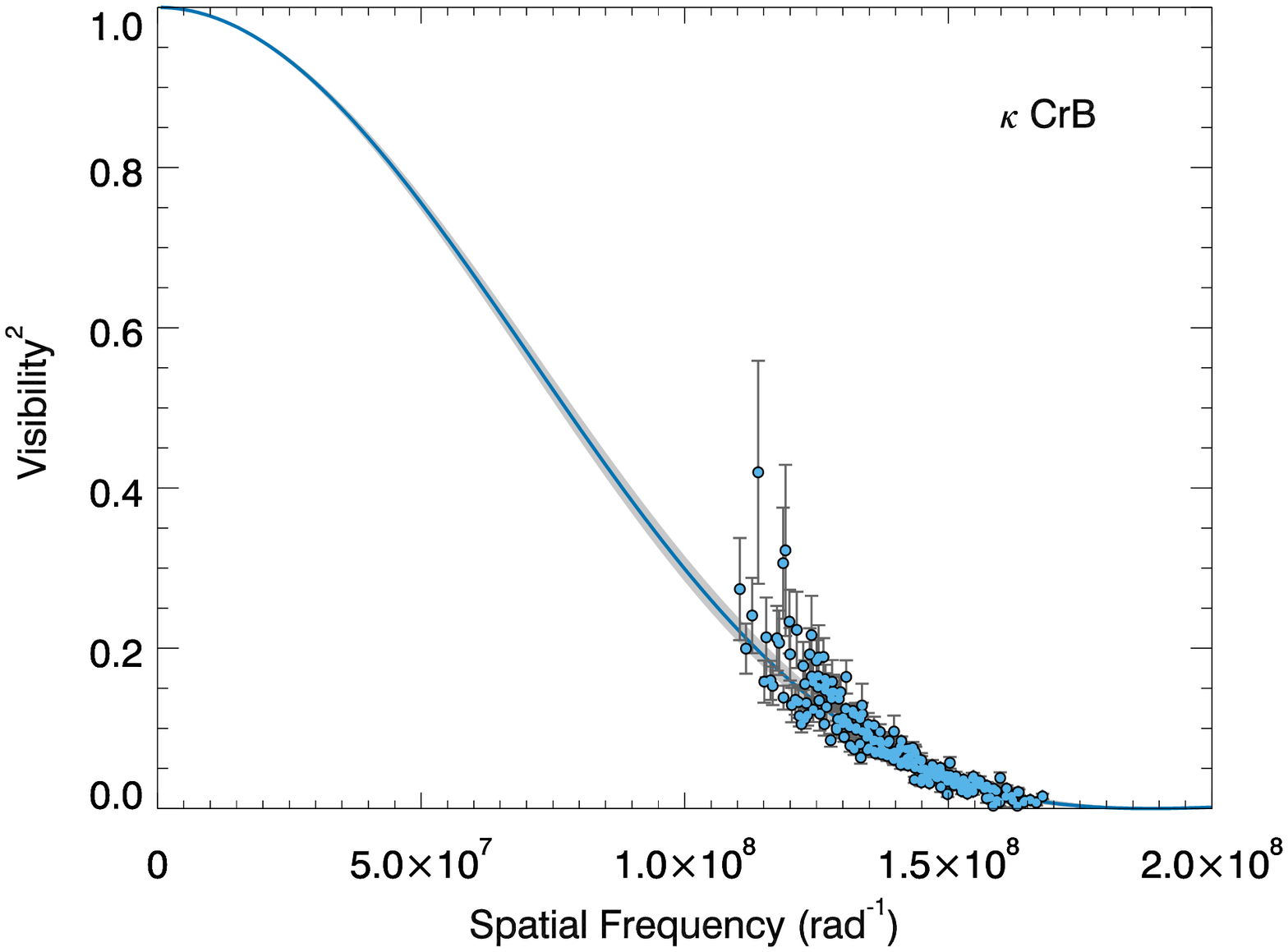}
    \caption{Squared visibility versus spatial frequency for \kcrb. The blue circles and line indicate the PAVO observations and best-fitting model, respectively, as for Fig.~\ref{fig:6Lyn}.}
    \label{fig:kapCrB}
\end{figure}   

\begin{table*}
\centering
\caption{Measured angular diameters, bolometric fluxes and fundamental properties}
\label{tab:newresults}
\begin{tabular}{lccccccccc}
\hline
Star & $u$ & $\theta_\mathrm{UD}$ & $\theta_\mathrm{LD}$ & $R$ & $F_\mathrm{bol, MARCS}$ & $F_\mathrm{bol, sp}$ & $T_\mathrm{eff}$ & $L$ \\
     &     & (mas)                & (mas)                & $\left(\mathrm{R}_\odot\right)$ & (pW.m$^{-2}$) & (pW.m$^{-2}$) & $\left(\mathrm{K}\right)$ & $\left(\mathrm{L}_\odot\right)$ \\
 \hline
6 Lyn        & 0.63$\pm$0.04 & 0.801$\pm$0.007 & 0.853$\pm$0.013 & 5.12$\pm$0.16 & 146.7$\pm$1.1 & 145.4$\pm$5.1 & 4949$\pm$58 & 14.2$\pm$0.9\\
24 Sex       & 0.63$\pm$0.04 & 0.617$\pm$0.005 & 0.659$\pm$0.009 & 5.49$\pm$0.18 &  80.7$\pm$1.2 &  84.0$\pm$3.8 & 4908$\pm$65 & 15.8$\pm$1.2\\
$\kappa$ CrB & 0.64$\pm$0.04 & 1.361$\pm$0.009 & 1.456$\pm$0.020 & 4.77$\pm$0.07 &   397$\pm$5   &   398$\pm$11  & 4870$\pm$47 & 11.6$\pm$0.3 \\
HR 6817      & 0.63$\pm$0.04 & 0.772$\pm$0.006 & 0.823$\pm$0.011 & 4.45$\pm$0.08 & 126.9$\pm$1.7 & 142.6$\pm$5.5 & 5013$\pm$59 & 10.0$\pm$0.3 \\
HR 8461      & 0.63$\pm$0.04 & 0.778$\pm$0.007 & 0.831$\pm$0.011 & 4.91$\pm$0.12 & 133.1$\pm$1.6 & 138.2$\pm$6.5 & 4950$\pm$68 & 13.1$\pm$0.8\\
\hline
\end{tabular}
\end{table*}

\begin{figure}
	\includegraphics[width=\columnwidth]{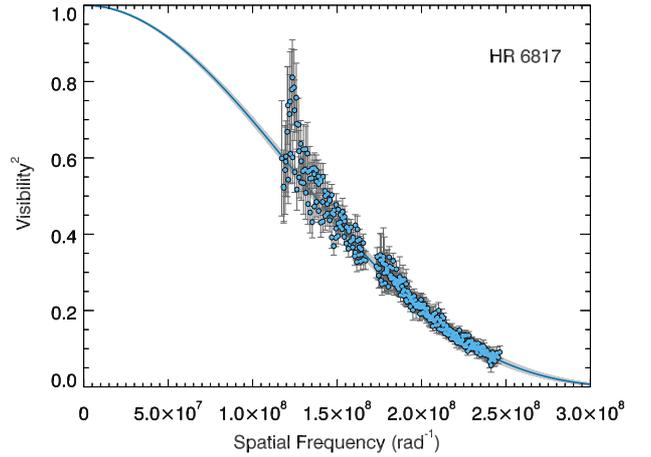}
    \caption{Squared visibility versus spatial frequency for \hdone. The blue circles and line indicate the PAVO observations and best-fitting model, respectively, as for Fig.~\ref{fig:6Lyn}.}
    \label{fig:HD167042}
\end{figure}

\begin{figure}
	\includegraphics[width=\columnwidth]{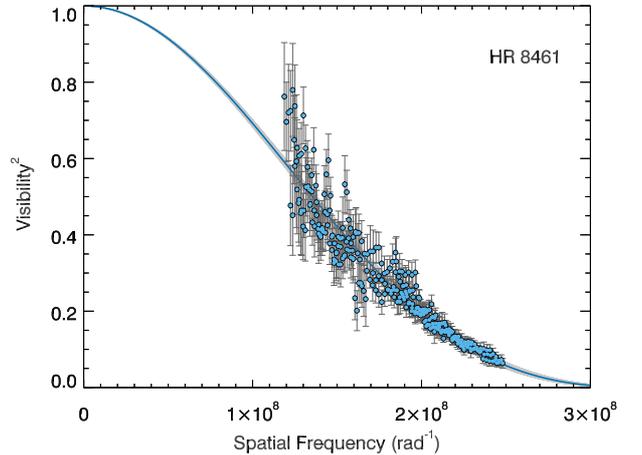}
    \caption{Squared visibility versus spatial frequency for \hdtwo. The blue circles and line indicate the PAVO observations and best-fitting model, respectively, as for Fig.~\ref{fig:6Lyn}.}
    \label{fig:HD210702}
\end{figure}

Figures~\ref{fig:6Lyn}--\ref{fig:HD210702} present the calibrated squared-visibility measurements as a function of spatial frequency (that is, the ratio of the projected baseline to the wavelength of the observation) of \lyn, \sex, \kcrb, \hdone, and \hdtwo, respectively. The calibrated fringe visibilities were fitted with a linearly limb-darkened disc model, given by \citep{hanburybrown74}
\begin{equation}
V = \left( \frac{1-u}{2} + \frac{u}{3} \right)^{-1}\left[ (1-u) \frac{J_1(x)}{x} + u (\pi/2)^{1/2} \frac{J_{3/2}(x)}{x^{3/2}} \right], \label{eqn:ldvis}
\end{equation}
where $x \equiv \pi B \theta_\mathrm{LD} \lambda^{-1}$, $V$ is the visibility, $u$ is the wavelength-dependent linear limb-darkening coefficient, $J_n(x)$ is the $n^\mathrm{th}$ order Bessel function of the first kind, $B$ is the projected baseline, and $\lambda$ is the wavelength at which the observations were made.

The linear limb-darkening coefficients were determined from the grids derived from model atmospheres by \citet{claret11} and \citet{magic15}. The grids were interpolated to spectroscopic values of \teff, \logg, and \feh\ found in the literature, and given in Table~\ref{tab:spect}. \citet{claret11} used two different methods to determine the limb-darkening coefficients from 1D \textsc{atlas} models. The first was a simple least-squares fit to the computed intensity distribution. Subsequent integration of this parameterized version of the intensity distribution will lead to the flux not being accurately recovered, so they also presented a limb-darkening coefficient from a flux-conserving method. For each star we consider here, the value from the flux conservation and least-squares methods were below and above the values determined from the model grid of \citet{magic15}, respectively. We therefore adopted the value from the grid of \citet{magic15}, which was derived from 3D hydrodynamical models created with the \textsc{stagger} code, and took the difference between the two values determined from the grids of \citet{claret11} to be indicative of the systematic uncertainty. These adopted values are given in Table~\ref{tab:newresults}.
		
Following the procedure outlined by \citet{derekas11}, the model-fitting and parameter uncertainty estimation was performed using Monte Carlo simulations that took into account uncertainties in the visibility measurements, adopted wavelength calibration (0.5 per cent), calibrator sizes (5 per cent) and limb-darkening coefficients. Combining the measured limb-darkened angular diameter with the Hipparcos parallax \citep{vanleeuwen07} gives the linear radii, while combining the angular diameter with the measured bolometric flux, $F_\mathrm{bol,sp}$, gives the effective temperature. All measured fundamental properties are given in Table~\ref{tab:newresults}.

\section{Discussion}

\subsection{Comparison with previous interferometric measurements}
\begin{figure}
	\includegraphics[width=\columnwidth]{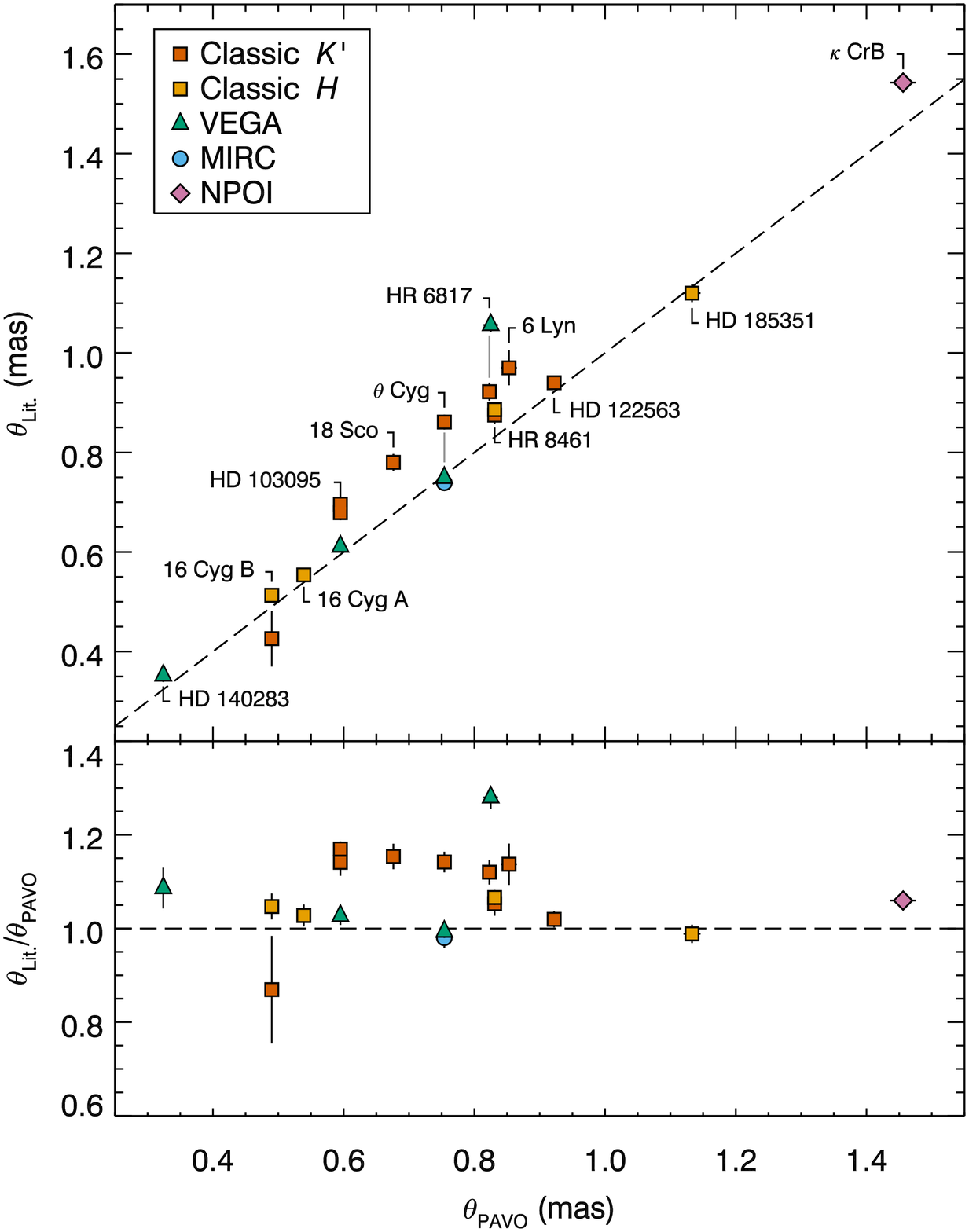}
    \caption{Comparison of CHARA Classic $K'$ and $H$ band (orange and yellow squares, respectively), VEGA (green triangles), MIRC (blue circle), and NPOI (pink diamond) measurements with PAVO measurements of the same stars.}
    \label{fig:pavoclascomp}
\end{figure}

Three of our targets -- \lyn, \hdone, and \hdtwo~-- have been previously observed with the CHARA Classic beam combiner. This adds to a sample of stars that has now been observed with both Classic and PAVO, a full list of which is given in Table~\ref{tab:compare}. Additionally, \kcrb\ has been previously observed with NPOI \citep[$\theta_\mathrm{LD}=1.543\pm0.009\,$mas;][]{Baines13}, while \hdone\ has also been observed with the VEGA beam combiner at the CHARA array \citep[$\theta_\mathrm{LD}=1.056\pm0.014\,$mas;][]{Ligi16}. Given the previously reported discrepancy between some photometric and interferometric temperatures \citep{Casagrande14}, it is worth considering how well these measurements made with different interferometric instruments compare, which we illustrate in Fig.~\ref{fig:pavoclascomp}. We consistently find Classic diameters that are systematically larger than those determined by PAVO. In some cases, the Classic diameters are 15 per cent larger than PAVO values, and disagreeing by up to 6$\sigma$. Notably, the largest differences are found for the Classic measurements made in $K'$ band.

\begin{table}
    \centering
    \caption{CHARA Classic versus PAVO angular diameters}
    \label{tab:compare}
    \begin{tabular}{lccccc}
        \hline
        Star & Classic $\theta_\mathrm{LD}$ & Band & Ref. & PAVO $\theta_\mathrm{LD}$ & Ref.\\
             & (mas)                        & &      & (mas)                  &        \\
        \hline
        16 Cyg B      & 0.426$\pm$0.056 & $K'$ & 1 & 0.490$\pm$0.006 & 9  \\
                      & 0.513$\pm$0.012 & $H$  & 2 &                 &    \\
        16 Cyg A      & 0.554$\pm$0.011 & $H$  & 2 & 0.539$\pm$0.006 & 9  \\
        HD 103095     & 0.696$\pm$0.005 & $K'$ & 3 & 0.595$\pm$0.007 & 10 \\
                      & 0.679$\pm$0.015$^a$ & $K'$ & 4 &                &    \\
        18 Sco        & 0.780$\pm$0.017 & $K'$ & 3 & 0.676$\pm$0.006 & 11  \\
        $\theta$\,Cyg &0.861$\pm$0.015 & $K'$ & 3 & 0.754$\pm$0.009 & 9  \\
        \hdone        & 0.922$\pm$0.018 & $K'$ & 5 & 0.823$\pm$0.011 & 12 \\
        \hdtwo        & 0.875$\pm$0.018 & $K'$ & 6 & 0.831$\pm$0.011 & 12 \\
                      & 0.886$\pm$0.006 & $H$, $J$ & 7 &                 &    \\
        \lyn          & 0.970$\pm$0.035 & $K'$ & 6 & 0.853$\pm$0.013 & 12 \\
        HD\,122563    & 0.940$\pm$0.011$^a$ & $K'$ & 4 & 0.926$\pm$0.011 & 10 \\
        HD\,185351    & 1.120$\pm$0.018 & $H$  & 8 & 1.133$\pm$0.013 & 8  \\
\hline
\end{tabular}
\flushleft References: (1)~\citet{Baines08}; (2)~\citet{Boyajian13}; (3)~\citet{Boyajian12}; (4)~\citet{Creevey12}; (5)~\citet{Baines10}; (6)~\citet{Baines09}; (7)~\citet{vonBraun14}; (8)~\citet{Johnson14}; (9)~\citet{White13}; (10)~\citet{karovicova18}; (11)~\citet{Bazot11}; (12)~this work.
\newline $^a$ Value also includes observations made with the FLUOR instrument at the CHARA Array.
\end{table}

Differences are also found with diameters measured with other beam combiners. The VEGA measurement of \hdone\ gives a diameter that is 28 per cent larger than found with PAVO, differing by 13$\sigma$. \citet{Ligi16} had noted that their VEGA measurement was discrepant with the earlier value determined with Classic by \citet{Baines10}, which is itself 12 per cent (4.6$\sigma$) larger than the PAVO result. 

Two other stars have PAVO, VEGA and Classic measurements reported in the literature: $\theta$\,Cyg and HD 103095. Additionally, HD\,140283 has been observed by VEGA and PAVO only. For $\theta$\,Cyg, the VEGA measurement \citep[$0.749\pm0.008\,$mas;][]{Ligi16} agrees well with the PAVO measurement \citep[$0.754\pm0.009\,$mas;][]{White13}, as well as with $H$-band measurements made with the MIRC beam combiner at the CHARA Array \citep[$0.739\pm0.015\,$mas;][]{White13}. Again, the $K'$-band measurement with Classic \citep[$0.861\pm0.015\,$mas;][]{Boyajian12} is larger. For HD 103095, there is only a 1.4$\sigma$ difference between the VEGA (0.611$\pm$0.009\,mas) and PAVO (0.595$\pm$0.007\,mas) values \citep{karovicova18}, both of which are substantially smaller than the value obtained from the FLUOR and Classic beam combiners \citep[$0.679\pm0.015\,$mas;][]{Creevey12}. The PAVO diameter of HD\,140283 \citep[$0.324\pm0.005\,$mas;][]{karovicova18} is 2$\sigma$ smaller than the VEGA measurement \citep[$0.353\pm0.013\,$mas;][]{Creevey15}.

The only star measured by both PAVO and NPOI to date is \kcrb. Once again, we find disagreement, with the NPOI diameter being 5.8 per cent larger than what we have obtained with PAVO, a 3.7$\sigma$ difference.

The source of these disagreements is not readily apparent. Accurate calibration of interferometric data is difficult, and there are several potential sources of systematic errors. \citet{Casagrande14}, for example, observed the disagreement in effective temperature increased with smaller angular diameters. Additionally, we find the disagreement with the diameters measured by Classic to be most apparent in $K'$ band, that is, in the longest wavelength band used. Both of these observations suggest that problems may be arising when targets are under-resolved.

It is instructive to consider how uncertainties propagate into the calibrated fringe visibility. The corrected visibility of the target object is given by
\begin{equation}
    V_\mathrm{obj,cor} = \frac{V_\mathrm{obj,obs}}{V_\mathrm{sys}},
\end{equation}
with the system visibility, 
\begin{equation}
V_\mathrm{sys} = \frac{V_\mathrm{cal,obs}}{V_\mathrm{cal,pred}},
\end{equation}
where $V_\mathrm{obj,obs}$ and $V_\mathrm{cal,obs}$ are the observed visibility measurements of the object and calibrator stars respectively, and $V_\mathrm{cal,pred}$ is the predicted visibility of the calibrator star in an ideal system. 

The first requirement for an accurate calibration is an accurate estimate of $V_\mathrm{cal,pred}$. Systematic errors in the predicted diameters of calibrator stars will result in biased calibrated visibilities. Such biases can be minimized by the careful choice of calibrator stars. The ideal calibrator is a nearby point source, of similar brightness and colour as the target. The ideal calibrator does not exist, so compromises are necessary. Provided a calibrator is small enough to be barely-resolved, errors in the predicted visibility should be negligible. 

A check of the calibrator stars used in the literature raises only a few problematic cases. The four calibrator stars used for the FLUOR observations of HD\,103095 and HD\,122563 would all have been partially resolved on the baselines used, being 0.84--0.98\,mas in size \citep{Creevey12}. In the particular case of HD\,103095, the calibrator stars are much more resolved than the target. Additionally, the calibrator stars for the VEGA observations of HD\,140283 \citep{Creevey15} are only slightly smaller than the target. It must be noted that HD\,140283 is a particularly difficult target to observe due to its relatively small angular size, and nearby stars that are bright and yet small enough to serve as calibrators are therefore rare.

In another case we have found large discrepancies in the assumed size of a calibrator star used multiple times throughout the literature. HD\,177003, a B2.5IV star, was used as the calibrator star for the VEGA observations of \hdone\, where a diameter of 0.130$\pm$0.009\,mas was adopted \citep{Ligi16}. A significantly larger diameter of 0.198$\pm$0.010\,mas was adopted for calibrations of $\theta$\,Cyg and 16\,Cyg\,A and B by \citet{White13} and for HD\,185351 by \citet{Johnson14}, while \citet{Jones15} used 0.156$\pm$0.016\,mas when calibrating observations of 16\,Lyr. However, even if the true diameter of HD\,177003 is substantially larger than the value adopted by \citet{Ligi16}, this cannot explain the overly large diameter found with VEGA for \hdone\ because adopting a larger diameter for the calibrator would result in an even larger diameter for the target. Additionally, \citet{ligi12,Ligi16} used HD\,177003 as a calibrator for the VEGA observations of $\theta$\,Cyg, and that measurement agrees with the value obtained with PAVO, despite the differences in the adopted calibrator sizes. This underlines how robust the calibration is to large uncertainties in calibrator stars sizes provided they are unresolved.

A second requirement for accurate calibration is that $V_\mathrm{cal,obs}$ remains a reliable indication of the system visibility throughout observations of the target. The system visibility varies both spatially and temporally, sometimes rapidly. Although efforts are made to observe calibrators as close in position and time to the targets as possible, this may not be sufficient when the atmosphere is less stable. Such changes in the system visibility can be overcome if a sufficiently large number of independent observations are made, which is why we have sought multiple observations over multiple nights, on several baselines, and with different calibrators. By contrast, the observations of \lyn\ and \hdtwo\ by \citet{Baines09} and \hdone\ by \citet{Baines10} were taken on a single night on a single baseline with a single calibrator, potentially making these observations more vulnerable to variations in the system visibility.

Finally, these systematic effects can be minimized when the target star is well-resolved because the absolute size of the system visibility correction is smaller. The heteroskedasticity of the data seen in Figures\, \ref{fig:6Lyn}--\ref{fig:HD210702}, with observations at lower visibilities having less variance, is the result of this. The discrepant Classic $K'$-band diameters have arisen from observations where the lowest squared-visibility measurement is $\gtrsim 0.4$. In this category are the observations of HD\,103095 \citep{Creevey12}, 18\,Sco and $\theta$\,Cyg \citep{Boyajian12}, \lyn\ and \hdtwo\ \citep{Baines09} and \hdone\ \citep{Baines08}. The NPOI measurements of \kcrb\ were also made at $V^2\gtrsim 0.4$ \citep{Baines13}. The VEGA measurements of \hdone\ are similarly dominated by observations with $V^2\gtrsim 0.4$, with highly uncertain measurements around $V^2\approx 0$ contributing little to the fit. 

A potential explanation for the disagreements, then, is that natural variations in the system visibility are being aggravated when targets have relatively high visibilities, with an insufficient number of observations to gain a representative sample of the true measurement uncertainty. If this is the case, additional observations, particularly at higher spatial frequencies, should lead to results converging for the different interferometric instruments. This explanation, however, does not adequately explain why the apparent systematic errors only tend to lead to diameters that are too large. Some studies are now combining data from multiple instruments, with good agreement found in several cases \citep[e.g][]{Johnson14,karovicova18}. Additional investigations are, however, warranted to further get to the root of disagreements when they occur. 

\subsection{On the temperature and mass scale of `retired A stars'}
Interferometric angular diameters and bolometric flux measurements form the basis of the empirical effective temperature scale \citep[e.g.][]{Code76,Boyajian13}. Those relatively few stars for which these measurements exist have become benchmarks for calibrating large spectroscopic surveys \citep[e.g.][]{Jofre14,heiter15}. The inconsistencies between interferometric results, discussed above, are therefore a cause for concern. For the suspected retired A stars in our study, the smaller PAVO diameters imply effective temperatures that are $\sim$200\,K hotter.

Less direct methods for determining effective temperatures that are independent of angular size may be useful in distinguishing between discrepant interferometric measurements. It was by comparing differences between interferometric and photometric temperatures as a function of angular diameter that \citet{Casagrande14} identified apparent systematic biases in some interferometric radii.

\begin{figure}
	\includegraphics[width=\columnwidth]{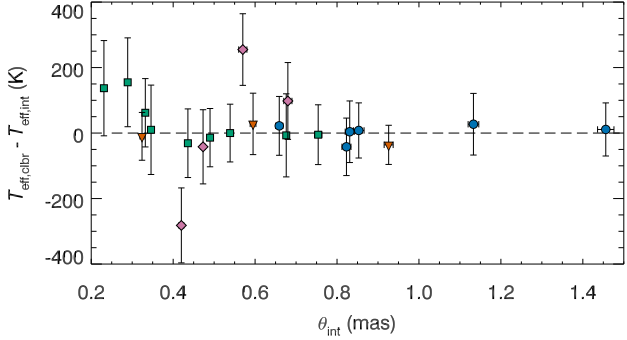}
    \caption{Difference between effective temperatures determined from a calibration of the IRFM and interferometry with PAVO. Retired A stars from this sample and HD\,185351 from \citet{Johnson14} are indicated by blue circles. Main-sequence stars from \citet{Bazot11}, \citet{huber12} and \citet{White13} are indicated by green squares, red giants from \citet{huber12} are indicated by pink diamonds, and metal-poor stars from \citet{karovicova18} are orange triangles.}
    \label{fig:tempcomp}
\end{figure}

In Fig.~\ref{fig:tempcomp} we compare our temperatures determined from PAVO interferometric measurements and those we determined using colour calibrations of Tycho2 photometry based on a sample for which the infrared flux method (IRFM) has been applied \citep{Casagrande06,Casagrande14}. With the possible exception of the two stars with $\theta < 0.3$\,mas, there is no evidence of a trend in the temperature difference as a function of angular diameter. It is possible that the two smallest stars may also be showing the effects of being under-resolved. Additionally, with only two red giant stars measured by \citet{huber12} having temperatures that disagree by more than 1$\sigma$, the negligible temperature differences show this calibration of the IFRM temperature scale is consistent with these interferometric measurements. The apparent underestimation of the uncertainties can be attributed to common systematics present in the absolute flux calibration
of photometric data.

\begin{figure}
	\includegraphics[width=\columnwidth]{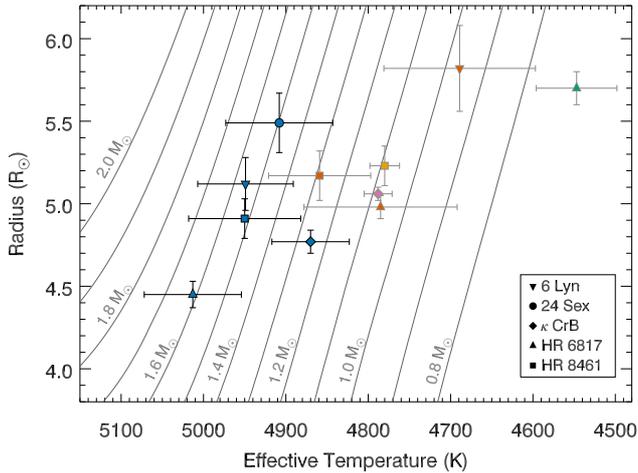}
    \caption{Radius--$T_\mathrm{eff}$ diagram for the suspected retired A stars in our sample, with each star identified by the symbol in the legend. Blue symbols indicate the values determined with PAVO in this work. Previous interferometric measurements are indicated with grey uncertainties, with the colour of the symbol indicating the beam combiner: CHARA Classic $K'$-band in orange and $H$-band in yellow, VEGA in green, and NPOI in pink. For reference, solar-metallicity BASTI evolutionary model tracks are shown in grey, from 0.8--2.0 M$_\odot$ as indicated.}
    \label{fig:radteff}
\end{figure}

The lack of a trend gives us confidence in the general accuracy of PAVO interferometric measurements of stars $\theta > 0.3$\,mas. We therefore conclude that the higher interferometric temperatures for the suspected retired A stars in our sample are accurate. With these stars located at the bottom of the red giant branch diagram, an increase in temperature corresponds to an increase in mass, as can be seen in Fig.~\ref{fig:radteff}.

Although the PAVO interferometric temperatures for these stars are significantly higher than previous interferometric determinations, they generally agree with the spectroscopic measurements given in Table~\ref{tab:spect}. Only the temperature of 24 Sex disagrees with the value found by \citet{Mortier13} by 1.8$\sigma$, with the interferometric temperature being cooler by 161$\pm$90\,K. We might therefore expect that the masses obtained using the interferometric measurements and constraints will generally agree with results in the literature.

\begin{table*}
    \centering
    \caption{Derived masses of the target stars from different sets of constraints}
    \label{tab:mass}
    \begin{tabular}{lcccccccc}
        \hline
        \multirow{3}{*}{Star} &  \multicolumn{8}{c}{Mass ($\mathrm{M}_\odot$)} \\
             & \multicolumn{2}{c}{Direct Method} & \multicolumn{6}{c}{Grid Modelling} \\ 
             &  &  &   \multicolumn{3}{c}{($V$, $\varpi$, $T_\mathrm{eff}^\mathrm{spec}$, [Fe/H])} & ($T_\mathrm{eff}^\mathrm{spec}$, [Fe/H], $\nu_\mathrm{max}$) & ($R_\mathrm{int}$, $T_\mathrm{eff}^\mathrm{int}$, [Fe/H]) & ($R_\mathrm{int}$, $T_\mathrm{eff}^\mathrm{int}$, $\nu_\mathrm{max}$)\\
        \hline
\lyn    & 1.37$\pm$0.22$^a$ & 1.44$\pm$0.23 &      ---      & 1.44$\pm$0.14$^c$ & 1.41$^{+0.06}_{-0.07}$ &  1.32$^{+0.17}_{-0.15}$ & 1.35$^{+0.12}_{-0.13}$ & 1.33$\pm$0.18 \\
\sex    &      ---      &      ---      & 1.81$\pm$0.08$^b$ &      ---      & 1.64$^{+0.15}_{-0.13}$ &      ---      & 1.30$^{+0.15}_{-0.13}$ & --- \\
\kcrb   & 1.40$\pm$0.21$^a$ & 1.44$\pm$0.22 & 1.58$\pm$0.08$^b$ & 1.50$^{+0.11}_{-0.12}$$^c$ & 1.32$\pm$0.10  & 1.26$\pm$0.14 & 1.29$\pm$0.11 & 1.37$^{+0.14}_{-0.18}$\\
\hdone  &      ---      &      ---      & 1.63$\pm$0.06$^b$ &      ---      & 1.45$^{+0.10}_{-0.12}$ &      ---      & 1.42$^{+0.11}_{-0.15}$ & --- \\
\hdtwo  & 1.47$\pm$0.23$^a$ & 1.61$\pm$0.25 & 1.71$\pm$0.06$^b$ & 1.61$^{+0.08}_{-0.09}$$^c$ & 1.53$^{+0.08}_{-0.09}$ & 1.33$^{+0.17}_{-0.16}$ & 1.37$^{+0.15}_{-0.13}$ & 1.43$^{+0.20}_{-0.18}$ \\
\hline
\end{tabular}
\flushleft $^a$~Literature values from \citet{Stello17} \\ $^b$~Literature values from \citet{Mortier13} \\ $^c$~Literature values from \citet{Brewer16}
\end{table*}

A direct determination of the mass can be derived through the application of the asteroseismic scaling relation for the frequency of maximum power, $\nu_\mathrm{max}$, \citep{brown91,kjeldsenbedding95}
\begin{equation}
    \frac{M}{\mathrm{M}_\odot} \approx \left(\frac{R}{\mathrm{R}_\odot}\right)^{2} \left(\frac{T_\mathrm{eff}}{\mathrm{T}_{\mathrm{eff},\odot}}\right)^{1/2}
    \left(\frac{\nu_\mathrm{max}}{\nu_{\mathrm{max},\odot}}\right).
\end{equation}
\citet{Stello17} measured $\nu_\mathrm{max}$ for three of our stars; to determine mass they combined this measurement with the spectroscopic temperature and a determination of luminosity. The luminosity was derived from the Hipparcos parallax, Tycho $V_T$ magnitude, and spectroscopic temperature using \textsc{isoclassify} \citep{Huber17}. We are able to determine the mass more directly from $\nu_\mathrm{max}$ and the interferometric radius and temperature. These `direct method' masses are given in the first two columns of Table~\ref{tab:mass}. The relatively large uncertainty in these values is largely a consequence of the conservative 15 per cent assumed uncertainty in $\nu_\mathrm{max}$ adopted by \citet{Stello17}, and they consequently agree within these uncertainties.

A drawback from the direct values is that they do not take into account the slower evolution of lower-mass stars, which are therefore more likely to be observed \citep[see e.g.][]{lloyd11}. A less-biased constraint on mass can therefore be provided with reference to stellar evolutionary models.

We have determined the masses of the stars with reference to a grid of evolutionary models using the Bayesian Stellar Algorithm \citep[BASTA;][]{SilvaAguirre15,SilvaAguirre17}. The grid, used recently by \citet{SilvaAguirre18}, was constructed from BaSTI isochrones \citep{Pietrinferni04} including convective core overshooting during the main sequence and no mass loss. Different sets of observational constraints may be used to determine the best fitting models. 

For a direct comparison with published values derived from spectroscopy by \citet{Mortier13} and \citet{Brewer16}, we have applied our grid with the same constraints, namely $V$ magnitude, Hipparcos parallax $\varpi$, and the spectroscopic $T_\mathrm{eff}$ and [Fe/H]. The published  mass values and our determinations from BASTA are given are columns 3--5 of Table~\ref{tab:mass}. The BASTA masses tend to be smaller than those determined by \citet{Mortier13}, with a difference larger than 1$\sigma$ for three of four stars. Better agreement is found with the masses of \citet{Brewer16}. The differences in these values may be attributed to the different models used, with \citet{Mortier13} using an earlier version of the \textsc{param} tool \citep{daSilva06} using \textsc{parsec} models \citep{Bressan12}, and \citet{Brewer16} using Yale-Yonsei isochrones \citep{Demarque04}.

We have also used BASTA to determine masses using a combination of spectroscopic ($T_\mathrm{eff}$, [Fe/H]) and asteroseismic ($\nu_\mathrm{max}$) constraints, interferometric ($R$, $T_\mathrm{eff}$) and spectroscopic ([Fe/H]) constraints, and interferometric ($R$, $T_\mathrm{eff}$) and asteroseismic constraints ($\nu_\mathrm{max}$). These values are given in columns 6--8 of Table~\ref{tab:mass}, respectively. These masses are in agreement with each other. They also agree with the BASTA masses found from $V$, $\varpi$, and spectroscopic $T_\mathrm{eff}$ and [Fe/H], with the exception of \sex, for which the lower interferometric temperature contributes to a lower mass determination. As expected, these masses are lower than those determined directly from the scaling relation because of the slower evolution of lower-mass stars.

The slightly lower BASTA masses tend to support the conclusion of \citet{Stello17} that previous mass determinations that largely relied on spectroscopic parameters are, on average, overestimated. \citet{north17} did not find any strong evidence for a systematic bias in their sample of `retired' A (and F) stars, but noted the scatter in published masses, larger than quoted uncertainties, complicates comparisons. They suggested that differences in masses may be attributed to different constraints being applied. However, that BASTA also supports lower masses when the same spectroscopic parameters are used as constraints, and the generally good agreement between spectroscopic and our interferometric temperatures suggests that important differences may also be attributed to the choice of stellar models and their included physics. 

\section{Conclusions}
We have measured the angular diameters and bolometric fluxes of five planet-hosting low-luminosity red giant stars, and hence determined their radii and effective temperatures.

Significant differences of up to $\sim$\,30 per cent are found with interferometric measurements of these and other stars made with different instruments. The stars in our sample are better resolved by our new measurements, and our effective temperatures agree well with photometric and spectroscopic determinations. We suggest that the comparatively lower angular resolution of the earlier measurements has left them vulnerable to calibration errors, particularly when there are few independent measurements. Further studies are warranted to better understand these systematic effects.

We determined the masses of these stars using BASTA for combinations of spectroscopic, interferometric, and asteroseismic constraints. Masses from the different constraints were consistent with each other, but tended to be $\sim$15 per cent lower than values found in the literature, even when the same observational constraints are used. This suggests that variations in stellar models and how they are combined with observational constraints to determine stellar properties have a significant impact on the derived masses of these stars.

Additional asteroseismic observations of these stars will provide further insight to the masses of these stars. In particular, the upcoming NASA TESS Mission \citep{ricker15} will provide the opportunity to significantly expand the number of low-luminosity red giants with detected solar-like oscillations that are bright enough to be followed-up with long-baseline optical interferometry. Measurements of the characteristic frequency spacing between oscillation modes of consecutive radial order ($\Delta\nu$) allow the stellar density to be determined with great precision. This will allow for detailed studies that test stellar models through a combination of interferometry, asteroseismology, and spectroscopy, to be expanded to a wider sample of stars beyond HD\,185351 \citep{Johnson14,Hjoerringgaard17}.

\section*{Acknowledgements}
This work is based upon observations obtained with the Georgia State University Center for High Angular Resolution Astronomy Array at Mount Wilson Observatory, and the University of Hawai\okina i 2.2-m telescope on Maunakea. The authors wish to recognize and acknowledge the very significant cultural role and reverence that the summit of Maunakea has always had within the indigenous Hawaiian community. We are most fortunate to have the opportunity to conduct observations from this mountain. The CHARA Array is supported by the National Science Foundation under Grants No. AST-1211929 and AST-1411654. Institutional support has been provided from the GSU College of Arts and Sciences and the GSU Office of the Vice President for Research and Economic Development.

Funding for the Stellar Astrophysics Centre is provided by The Danish National Research Foundation. The research was supported by the ASTERISK project (ASTERoseismic Investigations with SONG and Kepler) funded by the European Research Council (Grant agreement no.: 267864). TRW and VSA acknowledge the support of the Villum Foundation (research grant 10118). DH acknowledges support by the Australian Research Council's Discovery Projects funding scheme (project number DE140101364) and support by the NASA Grant NNX14AB92G issued through the \textit{Kepler} Participating Scientist Program. LC is supported by the Australian Research Council Future Fellowship FT160100402. MI was supported by the Australian Research Council Future Fellowship FT130100235.




\bibliographystyle{mnras}
\bibliography{references} 


\clearpage
\newpage
\appendix

\section{Additional Figures}\label{apx:fbol}
\begin{figure}
    \includegraphics[width=\columnwidth]{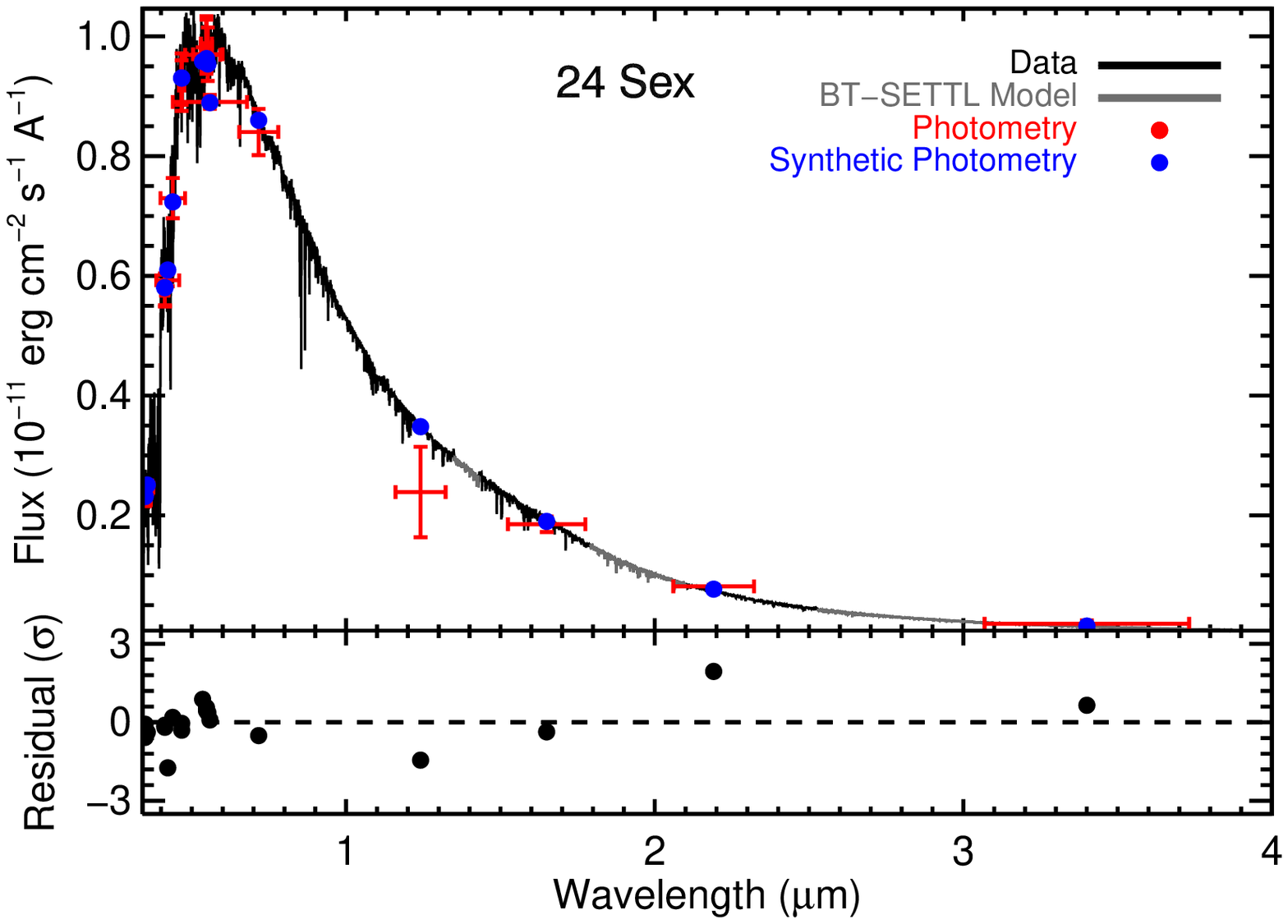}
    \caption{Absolutely calibrated spectrum of \sex, from which we compute \fbol. Lines and symbols as for Fig.~\ref{fig:fbol}.
    }
    \label{fig:fbol-24Sex}
\end{figure}

\begin{figure}
    \includegraphics[width=\columnwidth]{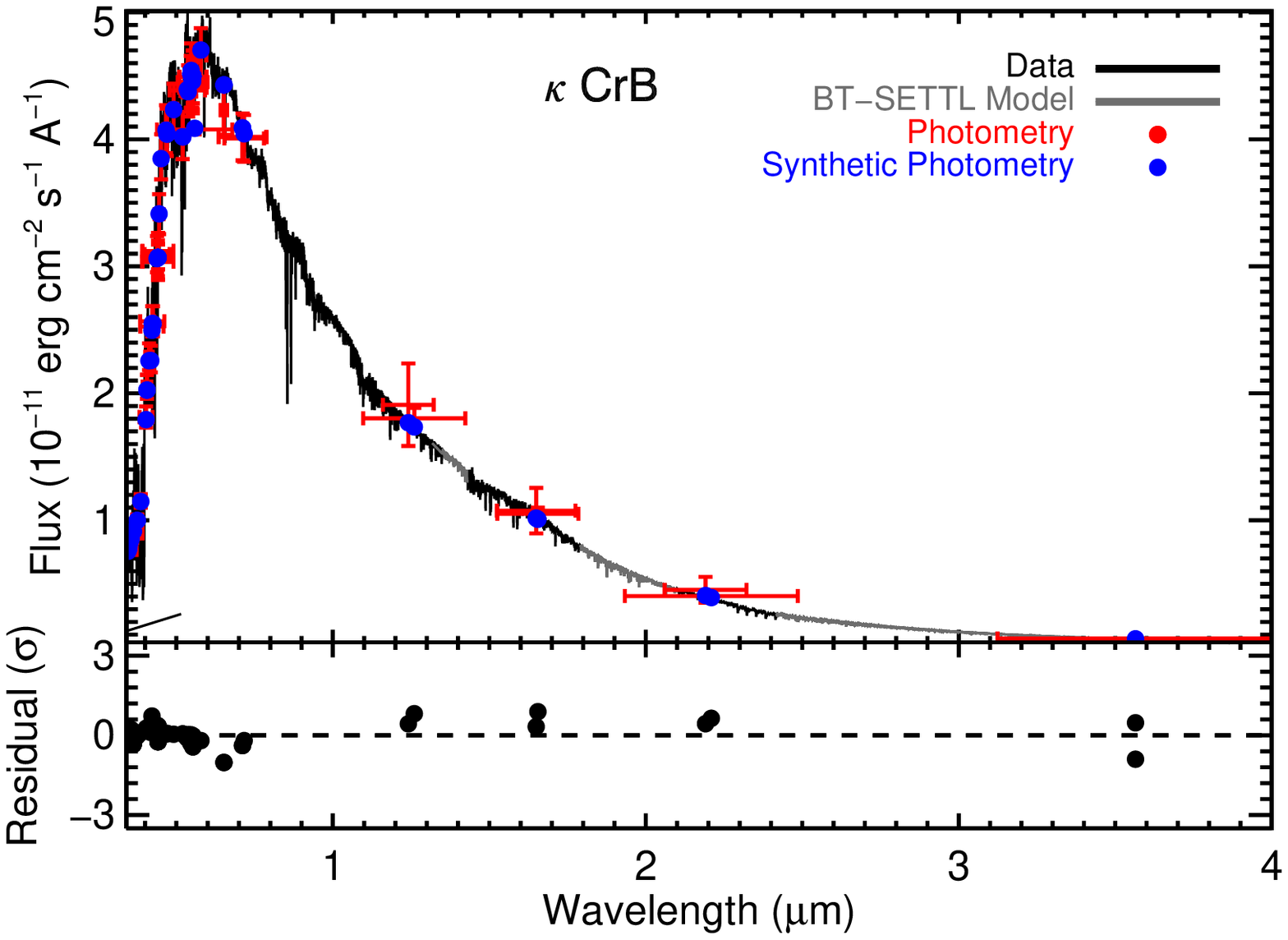}
    \caption{Absolutely calibrated spectrum of \kcrb, from which we compute \fbol. Lines and symbols as for Fig.~\ref{fig:fbol}, with the exception that the optical spectrum was obtained from Hubble's Next Generation Spectral Library \citep[NGSL;][]{Heap2007}, instead of SNIFS.
    }
    \label{fig:fbol-kapCrB}
\end{figure}

\begin{figure}
    \includegraphics[width=\columnwidth]{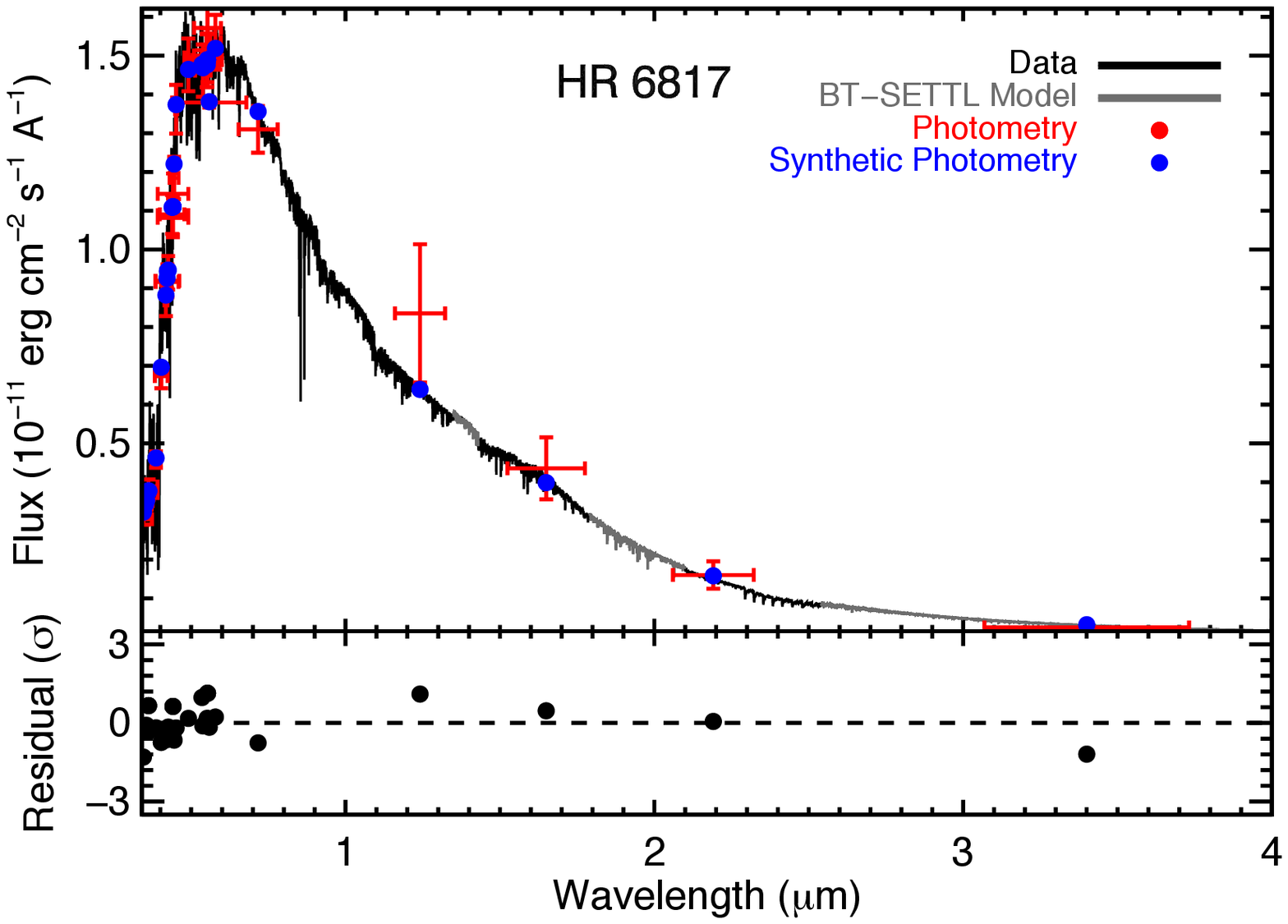}
    \caption{Absolutely calibrated spectrum of \hdone, from which we compute \fbol. Lines and symbols as for Fig.~\ref{fig:fbol}.
    }
    \label{fig:fbol-hr167042}
\end{figure}

\begin{figure}
    \includegraphics[width=\columnwidth]{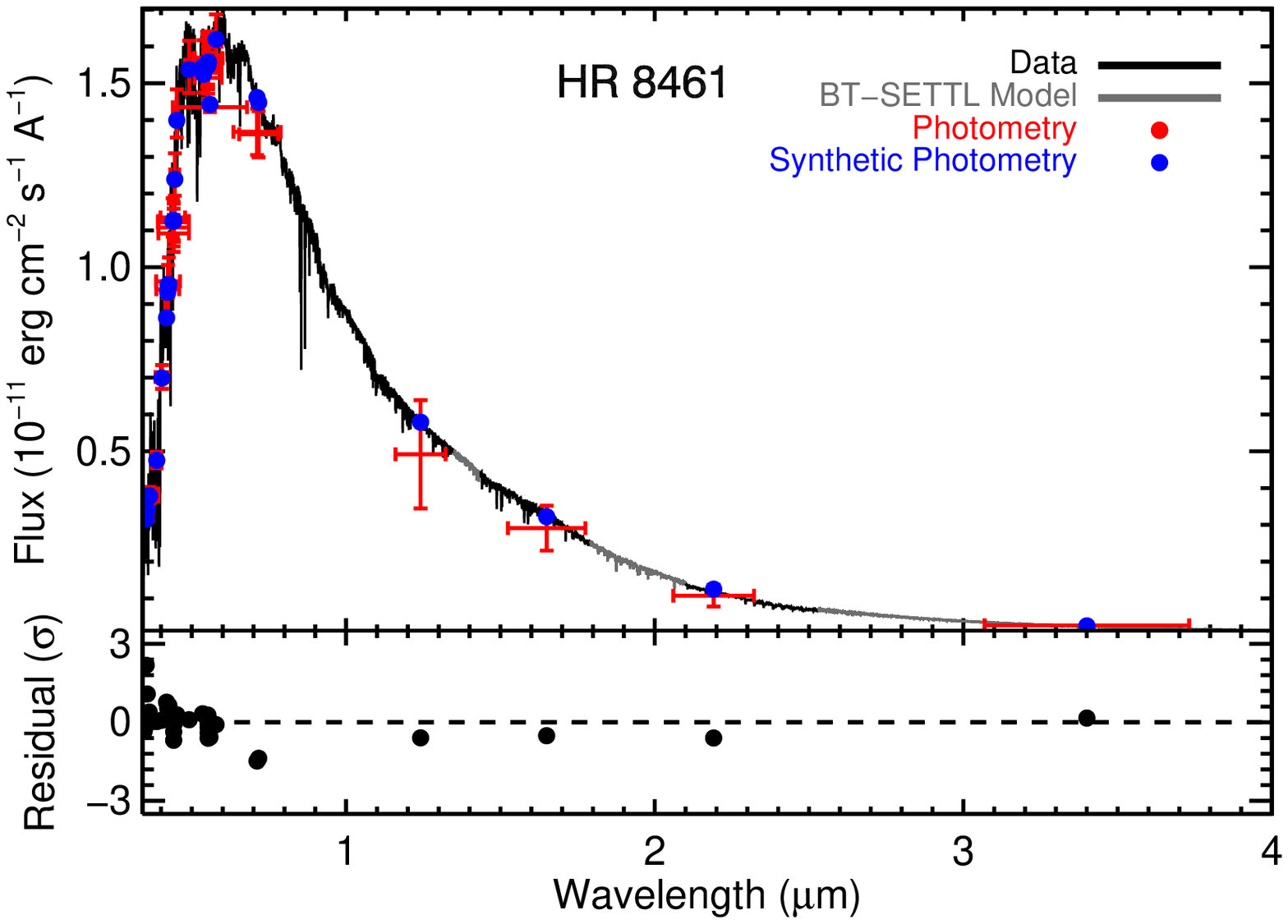}
    \caption{Absolutely calibrated spectrum of \hdtwo, from which we compute \fbol. Lines and symbols as for Fig.~\ref{fig:fbol}.
    }
    \label{fig:fbol-hr210702}
\end{figure}

Figures~\ref{fig:fbol-24Sex}--\ref{fig:fbol-hr210702} show the flux-calibrated spectra of \sex, \kcrb, \hdone, and \hdtwo, respectively.

\bsp	
\label{lastpage}
\end{document}